# Joint Beamforming and Power Allocation for MIMO Relay Broadcast Channel with Individual SINR Constraints

Chun-Che Chien, Hsuan-Jung Su, and Hsueh-Jyh Li

*Abstract*—In this paper, system design for the multi-input multi-output (MIMO) relay broadcast channel with individual signal-to-interference-plus-noise ratio (SINR) constraints at the mobile stations (MS) is considered. By exploring the structure of downlink (DL) uplink (UL) duality at either the base station (BS) or the relay station (RS), we propose two schemes of joint power allocation and beamforming design at the BS and the RS. The problem of existence of feasible solutions under practical power constraints at the BS and the RS with given SINR targets is considered first. Then the problem of sum power minimization is considered. Each design problem can be solved efficiently using optimal joint power allocation and beamforming under the framework of convex optimization. We also show that with subchannel pairing at the RS, the transmission power can be reduced by channel compensation at either hop. Finally, an extension to more general multi-hop applications is provided to further improve the power efficiency.

## I. INTRODUCTION

In the recent years, multi-input-multi-output (MIMO) systems have been adopted in many wireless applications, such as in point-to-point multiple-antenna communications [1]-[3] and cellular multiple-user communications [4]-[6]. It is well known that with the deployment of multiple antennas at wireless terminals, channel capacity enhancement and robustness against channel fading can be achieved. On the other hand, with the deployment of relay stations at cell edge or severely blocked area, the coverage area can be extended to enhance the throughput of cell-edge users. In order to exploit the above benefits, a promising compound scheme that incorporates MIMO technology into the relay architecture was introduced recently [7][11], where the base station (BS) and the relay station (RS) both have multiple antennas, and the mobile stations (MSs) are equipped with single antenna. Under such a configuration, a MIMO link between the BS and the RS followed by a multi-user downlink channel is established consequently. Half-duplex non-regenerative two-hop operation without direct link from the BS to the MSs is usually considered due to serious shadowing effect or the so-called dead zone position of the MSs. Various precoding schemes at the BS and the RS had been proposed under this structure in different applications. In [7], the achievable sum rate of the relay assisted MIMO downlink channel was derived assuming zero-forcing based dirty-paper-coding (DPC) at the BS and linear processing at the RS. In [8] and [9], the downlink-uplink duality under distributed single-antenna amplify-and-forward (AF) relays was established by assuming DPC based precoding at the BS. The precoding matrix in this scenario can be thought of as a diagonal matrix with joint RS operation. Although the DPC-based precoding promises the theoretically achievable data rate, in existing systems, the complexity and high cost issues restrict its prevalence. Recently, quality-of-service (QoS) based design criteria for multi-user relaying have been revealed. In [10], multiple single-antenna source nodes communicate with their corresponding destination nodes with the help of distributed single-antenna AF relays were considered. In that configuration, the joint precoding matrices at the source nodes and the relay nodes are both diagonal, and multi-user interferences can only be precluded with orthogonal channel relaying in the time or frequency domain. The QoS based MIMO relay broadcast channel, which is the multi-user broadcast channel with multi-antenna at the BS and the RS, can be solved by using the method of bi-convex optimization (that is, solving the convex problem at either the BS or the RS while assuming the design value at the other station is fixed) which iteratively updates the parameters until convergence is met [11]. Since the joint optimization problem for both the BS and the RS in the MIMO relay broadcast channel is non-convex, even though the bi-convex method optimizes the precoding matrices at the BS and the RS individually, it may result in a local optimum due to inadequate initial settings. An alternative scheme which is based on joint zero-forcing (ZF) design from the BS to the RS and from the RS to the MSs was also provided in [11]. Due to the effect of noise enhancement at both the RS and the MS, this scheme is inferior to the minimum mean square error (MMSE) based design.

In this paper, under the framework of MIMO relay broadcast channel similar to [7] and [11], we focus on the design of linear precoding at the BS and the RS in view of different QoS requirements for different MSs. In the QoS based design in [11], the total transmission power (for the BS and the RS) was minimized while the respective signal-to-interference-plus-noise-ratio (SINR) requirement at each MS was fulfilled. In practice, having individual transmission power constraints at the BS and the RS is more reasonable. Another important issue of the QoS based formulation is that certain design targets may not be achieved by any beamformer design and power allocation. To avoid endless search (iterations) of the algorithm, a criterion to judge if the target SINR set is feasible given BS and RS power constraints is required to help determine weather solving the subsequent power minimization problem makes sense.

In view of these design issues, in this paper, we first consider the feasibility of the system, then delve into the sum power minimization problem for the situations where feasible solutions exist. These problems are non-convex, and it is difficult to find the global optimal solution. We approach the problems by introducing two relaying structures, AF based relaying and singular value decomposition (SVD) based relaying, to divide the problem into two sub-problems: the power allocation problem at the BS and the RS, and the beamformer design problem at either the BS or the RS . The joint power allocation at the BS and the RS is identified as a geometric programming (GP) problem and converted to the problem of convex optimization. The beamformer design at the BS or the RS is conducted by exploring the existence of the downlink-uplink duality which transforms the difficult downlink problem into a more tractable uplink problem. Based on this, the corresponding uplink power allocation problem is also introduced for deriving the optimal transmit beamformer at the BS or the RS. We then provide iterative algorithms that can successively solve the above two optimization problems until convergence. Simulation results show that the proposed algorithm converges rapidly, and is thus computationally efficient for both feasibility test and the power minimization problem. In addition, subchannel pairing and generalized multi-hop relaying are proposed for the SVD based relaying design.

The rest of the paper is organized as follows. In Section II, a structure of the MIMO relay broadcast channel and the corresponding expressions of the signal flows are described. Section III contains the formulations of the optimization problems for feasibility test and power minimization. The proposed relaying structures which employ the theory of downlink-uplink duality are presented in Section IV for the AF based relaying and the SVD based relaying. In Section V, simulations are conducted to compare the performance of the proposed schemes and the existing works. Finally, Section VI concludes the paper.

The notational conventions used in this paper are: Vectors are in bold-face lower-case letters and matrices are in boldface upper-case letters. Subscripts $(\bullet)^T$ and $(\bullet)^H$ denote transpose and conjugate transpose, respectively. $\mathrm{Tr}(\bullet)$ denotes the trace operation and $diag(s_1,...,s_N)$ denotes a diagonal square matrix with $s_1,...,s_N$ denoting the diagonal elements. $\|\mathbf{s}\|$ denotes the Euclidean norm of the vector $\mathbf{s}$. $\mathbb{R}^+$ denotes the space of positive real numbers and $\mathbb{C}^{x \times y}$ denotes the space of $x \times y$ complex valued matrices. $\mathbf{I}_M$ denotes the identity matrix with size $M \times M$. We define $\mathbf{1} = [1,...,1]^T$. $Eig(\mathbf{R}_1, \mathbf{R}_2)$ denotes the generalized eigenvalue problem with $\mathbf{R}_1 \mathbf{x} = \lambda \mathbf{R}_2 \mathbf{x}$, where $\mathbf{x}$ and $\lambda$ are the eigenvector and eigenvalue, respectively. The collection set $\{1,...,K\} \setminus \{k\}$ indicates elements from 1 to $K$ except $k$.

## II. System Model

Consider a MIMO relay downlink channel depicted in Fig.1. The BS serves multiple MSs simultaneously with the help of a single RS. In this two-hop relay model, we neglect the direct link between the base station and the $K$ MSs. The BS and the RS are equipped with $M_b$ and $M_r$ antennas, respectively. All MSs under the coverage of the RS are single-antenna devices. A MIMO channel is formed between the BS and the RS. The BS first precodes the data that is targeted for multiple MSs and then sends it to the RS, the RS precodes the data using a transfer matrix before broadcasting the precoded signal to the MSs.

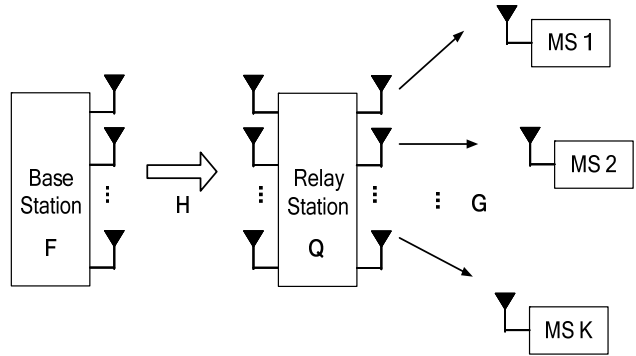

Fig .1. MIMO relay downlink channel.

The transmitted signal vector at the BS can be expressed as

$$\mathbf{x}_b = \mathbf{F}\mathbf{s} \tag{1}$$

where $\mathbf{s} = \left[ s_1, \cdots, s_K \right]^T$ denotes the transmitted signal vector intended for the $K$ MSs. $\mathbf{F} \in \mathbb{C}^{M_b \times K}$ is a precoding matrix composed of

$\mathbf{f}_k \in \mathbb{C}^{M_b \times 1}$, $k = 1, \ldots, K$, each denoting a transmit beamformer vector for MS $k$, i.e., $\mathbf{F} = \left[ \mathbf{f}_1, \cdots, \mathbf{f}_K \right]$. The received signal vector at

the RS can be expressed as

$$\mathbf{y}_r = \mathbf{H}\mathbf{x}_b + \mathbf{n}_r \tag{2}$$

where $\mathbf{H} \in \mathbb{C}^{M_r \times M_b}$ represents the MIMO channel between the BS and the RS, and $\mathbf{n}_r$ is the received noise vector at the RS. The elements in $\mathbf{n}_r$ are independent and identically distributed (i.i.d.) complex Gaussian random variables with zero mean and variance $\sigma_r^2$, i.e., $\mathbf{n}_r \sim CN(0, \sigma_r^2 \mathbf{I}_{M_r})$. If we use an $M_r \times M_r$ matrix $\mathbf{Q}$ to represent the linear precoding at the RS, the

re-transmitted signal vector at the output of RS is

$$\mathbf{x}_r = \mathbf{Q}\mathbf{y}_r. \tag{3}$$

With the definition of channel vector $\mathbf{g}_k$, $k = 1, \ldots, K$, of size $M_r \times 1$ between the RS and the $k$th MS, from (1), (2) and (3), the $k$th

MS observes the following combination of the transmitted signals from the BS

$$\begin{aligned} y_k &= \mathbf{g}_k^H \mathbf{Q} \left( \mathbf{H}\mathbf{F}\mathbf{s} + \mathbf{n}_r \right) + n_k \\ &= \mathbf{g}_k^H \mathbf{Q}\mathbf{H}\mathbf{f}_k s_k + \left( \mathbf{g}_k^H \mathbf{Q}\mathbf{H} \sum_{i=1, \neq k}^{K} \mathbf{f}_i s_i + \mathbf{g}_k^H \mathbf{Q}\mathbf{n}_r + n_k \right) \end{aligned} \tag{4}$$

where $n_k$ denotes the received noise at the $k$th MS, which is assumed to be an i.i.d. complex Gaussian random variable with zero mean and variance $\sigma_k^2$. On the right hand side of (4), the first term denotes the desired signal, and the overall inter-stream interferences, amplified noises and the local noise received at the $k$th MS are collected in the second term. If the input signal elements in vector $\mathbf{s}$ are assumed to be mutually independent and normalized to unit variance, the average transmit power at BS can

be derived from (1) as

$$P_b = \text{Tr}\left(\mathbf{x}_b \mathbf{x}_b^H\right) = \text{Tr}\left(\mathbf{F}\mathbf{F}^H\right) \tag{5}$$

Similarly, from (1) to (3), the average transmit power at RS can be shown as

$$P_r = \text{Tr}\left\{\mathbf{Q}\left(\mathbf{H}\mathbf{F}\mathbf{F}^H\mathbf{H}^H + \sigma_r^2\mathbf{I}\right)\mathbf{Q}^H\right\}. \tag{6}$$

From (4), the total power of received interference and noise at the $k$th MS is

$$\begin{aligned}
\tilde{\sigma}_k^2 &= \mathbf{g}_k^H\mathbf{Q}\mathbf{H}\left(\sum_{i=1,\neq k}^K \mathbf{f}_i\mathbf{f}_i^H\right)\mathbf{H}^H\mathbf{Q}^H\mathbf{g}_k + \left(\mathbf{g}_k^H\mathbf{Q}\mathbf{Q}^H\mathbf{g}_k\right)\sigma_r^2 + \sigma_k^2 \\
&= \sum_{i=1,\neq k}^K \left|\mathbf{g}_k^H\mathbf{Q}\mathbf{H}\mathbf{f}_i\right|^2 + \sigma_r^2\left\|\mathbf{g}_k^H\mathbf{Q}\right\|^2 + \sigma_k^2
\end{aligned}. \tag{7}$$

Thus, we have the SINR expression observed at the $k$t$h$ MS as

$$SINR_k = \frac{\left|\mathbf{g}_k^H\mathbf{Q}\mathbf{H}\mathbf{f}_k\right|^2}{\sum_{i=1,\neq k}^K \left|\mathbf{g}_k^H\mathbf{Q}\mathbf{H}\mathbf{f}_i\right|^2 + \sigma_r^2\left\|\mathbf{g}_k^H\mathbf{Q}\right\|^2 + \sigma_k^2}, \ k \in \{1,...,K\}. \tag{8}$$

For notational convenience, let $\mathbf{G} = [\mathbf{g}_1,...,\mathbf{g}_K]$. If the perfect channel knowledge of $\mathbf{H}$ and $\mathbf{G}$ is available at the RS, the precoders, $\mathbf{F}$ and $\mathbf{Q}$, for the BS and the RS, respectively, can be jointly determined at the RS in advance. The RS can then feedback $\mathbf{F}$ to the BS to fulfill the precoded two-hop transmission under individual SINR requirements of the MSs. It should be noticed that the power consumptions at the BS and the MS are implicitly determined by the precoding matrices $\mathbf{F}$ and $\mathbf{Q}$ according to equations (5) and (6) respectively.

### III.  Problem Formulation

In this section, the optimization problem for MIMO relay downlink channel is formulated. Instead of the maximum sum SINR based design criterion [7], which only favors users with better channel qualities, individual SINR constraints for the MSs are considered to emphasize the fairness in QoS. Assume that the MSs  have their individual SINR targets, $\gamma_1,...,\gamma_K$. To fulfill the target SINRs of all MSs simultaneously, the following condition has to be satisfied

$$\min_{k=1,...,K} \frac{SINR_k}{\gamma_k} \geq 1. \tag{9}$$

For power efficiency, we aim at joint design of $\mathbf{F}$ and $\mathbf{Q}$ to minimize the sum power of the BS and the RS under the constraint that the set of target SINRs can be satisfied. In addition, the individual power consumptions at the BS and the RS cannot exceed their corresponding maximum values. Therefore, we have the power minimization problem formulated as:

$$\underset{\mathbf{F},\mathbf{Q}}{\text{minimize}}\left(P_b + P_r\right)$$
$$subject \text{ to } \text{SINR}_k \geq \gamma_k, \ k = 1,...,K$$
$$P_b \leq P_b^{\max}$$
$$P_r \leq P_r^{\max}$$
(10)

where $P_b^{\max}$ and $P_r^{\max}$ are the maximum available power at the BS and the RS, respectively. We focus on the station-wise maximum power constraints since in most applications, the BS and the RS have separate power sources. Due to the individual maximum power constraints, the optimization problem (10) does not always have a solution. We first focus on the feasibility of the optimization problem using the following test problem.

$$\underset{\mathbf{F},\mathbf{Q}}{\max} \ \underset{i=1,...,K}{\min} \left(\frac{\text{SINR}_k}{\gamma_k}\right)$$
$$subject \text{ to } P_b \leq P_b^{\max}$$
$$P_r \leq P_r^{\max}$$
(11)

The target SINR set is feasible if and only if the objective function in (11) is greater than or equal to one. If the objective functions in both problems (10) and (11) were concave, then their global optima could be easily solved. Unfortunately, this is not the case when $\mathbf{F}$ and $\mathbf{Q}$ are considered jointly. Thus there is no efficient way to find the global optimal solution. To work around this situation, in the following sections, we introduce two design structures of $\mathbf{F}$ and $\mathbf{Q}$, which facilitate solving the problems via the downlink-uplink duality at the BS or the RS for iteratively computing the power allocation at the BS and the RS and the corresponding beamformer designs.

## IV. Design of Power Allocation and Beamformers at the BS and the RS

### A. Amplify and Forward Relay with Downlink-Uplink Duality at the BS

In this section, a simple amplify-and-forward (AF) relaying structure, where the RS only amplifies and forwards the received signal vector using a specific gain factor without linear processing, is proposed. In this structure, the precoding matrix at the RS is constructed as a scaled identity matrix, i.e., $\mathbf{Q} = \mathbf{\Lambda}_r^{1/2} = g_r^{1/2}\mathbf{I}_{M_r}$, where $g_r$ ($g_r \in \mathbb{R}^+$) represents the power amplifying factor at the RS, and $\mathbf{I}_{M_r}$ is an identity matrix of size $M_r \times M_r$. The equivalent channel matrix of size $K \times M_b$ from the BS to the MSs is then

$$\mathbf{H}_{eq} = \mathbf{G}^H \mathbf{\Lambda}_r^{1/2} \mathbf{H} .$$
(12)

This MIMO relay broadcast channel can be re-modeled as an equivalent MIMO broadcast channel with noise power at the $k$th MS as $\sigma_r^2 g_r \left\| \mathbf{g}_k^H \right\|^2 + \sigma_k^2$. Let $\mathbf{FF}^H = \mathbf{W}\mathbf{\Lambda}_p\mathbf{W}^H$, where $\mathbf{W} = [\mathbf{w}_1,...,\mathbf{w}_K]$, $\left\| \mathbf{w}_k \right\|^2 = 1$, $k = 1,...,K$, is a normalized precoder matrix consisting of unit norm beamformers, $\mathbf{\Lambda}_p = diag\left(p_1,...,p_K\right)$ is a diagonal matrix constructed from the power allocation vector $\mathbf{p} = [p_1, ..., p_K]^T$ at the BS. The formulation of the feasibility test problem (11) can be expressed as

$$
\begin{aligned}
&\max_{\mathbf{W}, \mathbf{\Lambda}_p, \mathbf{\Lambda}_r} \min_{i=1,...,K} \frac{p_k \left| \mathbf{g}_k^H \mathbf{\Lambda}_r^{1/2} \mathbf{H} \mathbf{w}_k \right|^2}{\gamma_k \left( \sum_{i=1, \neq k}^{K} p_i \left| \mathbf{g}_k^H \mathbf{\Lambda}_r^{1/2} \mathbf{H} \mathbf{w}_i \right|^2 + \sigma_r^2 \left\| \mathbf{g}_k^H \mathbf{\Lambda}_r^{1/2} \right\|^2 + \sigma_k^2 \right)} \\
&subject\ to \\
&\quad \mathrm{Tr}\left\{ \mathbf{\Lambda}_p \right\} \leq P_b^{\max} \\
&\quad \mathrm{Tr}\left\{ \mathbf{\Lambda}_r^{1/2} \left( \mathbf{H} \mathbf{W} \mathbf{\Lambda}_p \mathbf{W}^H \mathbf{H}^H + \sigma_r^2 \mathbf{I} \right) \left( \mathbf{\Lambda}_r^H \right)^{1/2} \right\} \leq P_r^{\max}
\end{aligned}
\tag{13}
$$

Although the above optimization problem is still not a joint convex problem for all design parameters $\mathbf{W}, \mathbf{\Lambda}_p$ and $\mathbf{\Lambda}_r$, with the decomposition of $\mathbf{F}$ into $\mathbf{W}$ and $\mathbf{\Lambda}_p$, an iterative solution based on the downlink-uplink duality can be applied. Our strategy is to iterate between power allocation (to obtain $\mathbf{\Lambda}_p$, $\mathbf{\Lambda}_r$) and beamfromer optimization (to obtain $\mathbf{W}$) until the solution converges. Note that the formulation (13) is equivalent to a MIMO broadcast channel without relay. Thus the analysis in [12] can be used to show the convergence of the algorithm.

*1) Beamfomer Design via Downlink-Uplink Duality*

In this subsection, we consider the optimization of the precoding matrix at the BS, $\mathbf{W}$, by assuming that the power allocations at the BS and the RS are fixed as $\mathbf{\Lambda}_p = \tilde{\mathbf{\Lambda}}_p = diag\left\{ \tilde{p}_1, ..., \tilde{p}_K \right\}$ and $\mathbf{\Lambda}_r = \tilde{\mathbf{\Lambda}}_r = \tilde{g}_r \mathbf{I}_{M_r}$. Under this assumption, from (8), the received downlink SINR for the $k$th MS can be written as:

$$
SINR_k^D \left( \mathbf{W}, \tilde{\mathbf{\Lambda}}_p, \tilde{\mathbf{\Lambda}}_r \right) = \frac{\left| \tilde{p}_k^{1/2} \tilde{\mathbf{h}}_k^H \mathbf{w}_k \right|^2}{\sum_{i=1, \neq k}^{K} \left| \tilde{p}_i^{1/2} \tilde{\mathbf{h}}_k^H \mathbf{w}_i \right|^2 + \hat{\sigma}_k^2}
\tag{14}
$$

where $\tilde{\mathbf{h}}_k^H = \mathbf{g}_k^H \tilde{\mathbf{\Lambda}}_r^{1/2} \mathbf{H}$ and $\hat{\sigma}_k^2 = \sigma_r^2 \left\| \mathbf{g}_k^H \tilde{\mathbf{\Lambda}}_r^{1/2} \right\|^2 + \sigma_k^2$. The downlink SINR of the $k$th MS in (14) is a function of $\mathbf{w}_k$ as well as the other beamformers $\mathbf{w}_i$, and it is difficult to determine all the optimum transmit beamformers at once. Observing that the expression in (14) is equivalent to that of a MIMO broadcast channel without relay, with the received noise power $\hat{\sigma}_k^2$ for the $k$th MS, the downlink-uplink duality originally derived for the MIMO broadcast channel [12] can be applied here with some modifications. The downlink-uplink duality states that the achieved SINR region of the virtual uplink channel is equal to the achieved SINR region of the downlink channel by the same set of beamformers and the same total power constraint. The SINR of the virtual uplink channel corresponding to (14) can be expressed as

$$
SINR_k^U \left( \mathbf{w}_k, \tilde{\mathbf{\Lambda}}_q, \tilde{\mathbf{\Lambda}}_r \right) = \frac{\left| \tilde{q}_k \mathbf{w}_k^H \tilde{\mathbf{h}}_k \right|^2}{\sum_{i=1, \neq k}^{K} \left| \tilde{q}_i \mathbf{w}_k^H \tilde{\mathbf{h}}_i \right|^2 + \hat{\sigma}_k^2}
\tag{15}
$$

where $\tilde{\mathbf{\Lambda}}_q = diag\{\tilde{q}_1,...,\tilde{q}_K\}$ is a diagonal matrix consisting of the virtual uplink transmit power of the MSs. For a given $\tilde{\mathbf{\Lambda}}_q$ (which will be optimized later by a separate step), the virtual uplink SINR of user $k$ in (15) is only a function of the beamformer $\mathbf{w}_k$, which makes it possible for individual optimization.

It has been shown in [12], that with equal noise power, both downlink and uplink have the same SINR achievable regions under the same power constraints, i.e., at their respective optima, $\frac{SINR_k^D}{\gamma_k} = \frac{SINR_k^U}{\gamma_k}$, $k = 1,...,K$. In our case, $\hat{\sigma}_1^2 \neq \hat{\sigma}_2^2 = ... \neq \hat{\sigma}_K^2$. We can normalize the channels by their corresponding noise standard deviations to create a "normalized" downlink channel with $\tilde{\mathbf{h}}_k' = \tilde{\mathbf{h}}_k / \hat{\sigma}_k$ and unit noise variance. Then the results of [12][13] can be applied to obtain the optimal solution from the corresponding "normalized" virtual uplink.

The normalized virtual uplink SINR of the $k$th stream in (15) can be rewritten as:

$$SINR_k^U = \frac{\mathbf{w}_k^H \mathbf{R}_{s_k} \mathbf{w}_k}{\mathbf{w}_k^H \mathbf{R}_{n_k} \mathbf{w}_k}, \ k = 1, \cdots, K \tag{16}$$

where $\mathbf{R}_{s_k} = \frac{\tilde{q}_k}{\hat{\sigma}_k^2} \tilde{\mathbf{h}}_k \tilde{\mathbf{h}}_k^H$ and $\mathbf{R}_{n_k} = \sum_{i=1,\neq k}^{K} \frac{\tilde{q}_i}{\hat{\sigma}_i^2} \tilde{\mathbf{h}}_i \tilde{\mathbf{h}}_i^H + \mathbf{I}_k$ are the desired signal covariance matrix and the undesired signal covariance matrix of the $k$th stream respectively. By defining $\lambda = SINR_k^U$, the optimum solution of the uplink receive beamformer $\mathbf{w}_k$, $\mathbf{w}_k^{opt}$, is the dominant eigenvector of the generalized eigenvalue problem $\mathbf{R}_{s_k} \mathbf{w}_k = \lambda \mathbf{R}_{n_k} \mathbf{w}_k$. That is,

$$\mathbf{w}_k^{opt} = \arg \max_{\lambda} Eig\left(\mathbf{R}_{s_k}, \mathbf{R}_{n_k}\right) \tag{17}$$

where $Eig\left(\mathbf{R}_{s_k}, \mathbf{R}_{n_k}\right)$ is the generalized eigenvalue problem solver. The above optimal beamformer derivation for each data stream can be applied in either the feasibility test problem or the sum power minimization problem.

### 2) Power Allocation at BS and RS for the Feasibility Test Problem

Next, we consider the optimization of power allocations at the BS and the RS ($\mathbf{\Lambda}_p$, $\mathbf{\Lambda}_r$) by assuming that the precoding matrix $\mathbf{W}$ at the BS is fixed, i.e., $\mathbf{W} = \tilde{\mathbf{W}}$. For the feasibility test problem in (13), with given beamformers, from (14) and (15), the downlink SINR and virtual uplink SINR expressions for the $k$th MS can be written as

$$SINR_k^D = \frac{p_k g_r \left|\hat{h}_{k,k}\right|^2}{\sum_{i=1,\neq k}^{K} p_i g_r \left|\hat{h}_{k,i}\right|^2 + g_r \sigma_r^2 \left\|\mathbf{g}_k^H\right\|^2 + \sigma_k^2} \tag{18}$$

and

$$SINR_k^U = \frac{q_k g_r \left|\hat{h}_{k,k}\right|^2}{\displaystyle\sum_{i=1,\neq k}^{K} q_i g_r \left|\hat{h}_{i,k}\right|^2 + g_r \sigma_r^2 \left\|\mathbf{g}_k^H\right\|^2 + \sigma_k^2} \tag{19}$$

respectively, where we use the notation $\hat{h}_{x,y}$ to represent a scalar channel defined as $\hat{h}_{x,y} = \mathbf{g}_x^H \mathbf{H} \tilde{\mathbf{w}}_y$, $1 \leq x, y \leq K$.

### a)  Downlink Power Allocation

Based on (18), the feasibility test problem in (11) now can be expressed as:

$$\max_{p_1,\ldots,p_K,g_r} \min_{i=1,\ldots,K} \frac{p_k g_r \left|\hat{h}_{k,k}\right|^2}{\gamma_k \left(\displaystyle\sum_{i=1,\neq k}^{K} p_i g_r \left|\hat{h}_{i,k}\right|^2 + g_r \sigma_r^2 \left\|\mathbf{g}_k^H\right\|^2 + \sigma_k^2\right)}$$

$$s.t. \sum_{k=1}^{K} p_k \leq P_b^{\max}$$

$$\sum_{k=1}^{K}\sum_{j=1}^{K} g_r p_k \left|\mathbf{H}_r(j,k)\right|^2 + K g_r \sigma_r^2 \leq P_r^{\max} \tag{20}$$

where $\mathbf{H}_r = \mathbf{H}\tilde{\mathbf{W}}$ is the equivalent channel matrix from the signal source vector to the RS. The formulation in (20) appears to be also a non-convex problem. However, the inverse of the objective function has a mathematical form of a posynomial. By defining a design variable $t$ to convert the original problem to the mathematical expression of an epigraph form, we have the equivalent expression of (20) as

$$\min_{p_1,\ldots,p_K,g_r,t} t$$

$$s.t. \frac{\gamma_k \left(\displaystyle\sum_{i=1,\neq k}^{K} p_i g_r \left|\hat{h}_{k,i}\right|^2 + g_r \sigma_r^2 \left\|\mathbf{g}_k^H\right\|^2 + \sigma_k^2\right)}{p_k g_r \left|\hat{h}_{k,k}\right|^2} \leq t, \; k = 1,\ldots,K$$

$$\sum_{k=1}^{K} p_k \leq P_b^{\max}$$

$$\sum_{k=1}^{K}\sum_{j=1}^{K} g_r p_k \left|\mathbf{H}_r(j,k)\right|^2 + K g_r \sigma_r^2 \leq P_r^{\max} \tag{21}$$

Since the objective function in (21) is a monomial and the constraints are all posynomials, the design problem in (21) is a GP problem and can be further converted to a convex problem after logarithmic changes of design variables [14]. The problem can then be solved using, for example, interior-point algorithms [15].

### b)  Uplink Power Allocation

Here we consider the virtual uplink power allocation among the MSs in the feasibility test problem with the same total power constraint $P_b^{\max}$. The RS power amplifying factor $g_r$ is assumed to have been determined in the downlink power allocation problem with the corresponding RS power constraint in (20), and is assumed to be fixed here, i.e., $g_r = \tilde{g}_r$. Then, based on (19), the uplink power allocation problem for the feasibility test problem is

$$\max_{q_1,\dots,q_K} \min_{i=1,\dots,K} \frac{q_k \tilde{g}_r \left|\hat{h}_{k,k}\right|^2}{\gamma_k \left( \sum_{i=1,i\neq k}^{K} q_i \tilde{g}_r \left|\hat{h}_{i,k}\right|^2 + \hat{\sigma}_k^2 \right)}$$

$$s.t. \sum_{k=1}^{K} q_k \leq P_b^{\max}$$

(22)

where $\hat{\sigma}_k^2 = \tilde{g}_r \sigma_r^2 \left\|\mathbf{g}_k^H\right\|^2 + \sigma_k^2$. Due to the strictly monotonically increasing of the objective function in $q_k$ and monotonically decreasing of the objective function in $q_i$ for $i \neq k$, for the optimal solution of (22), the users shall achieve the same balanced level which is defined as $C^U = SINR_k^U / \gamma_k$ for $k = 1,\dots,K$. We can then transform the problem (22) into the following equation.

$$\mathbf{q}\frac{1}{C^U} = (\mathbf{D\Psi}^T)\mathbf{q} + \mathbf{D\sigma} \quad.$$

(23)

Where $\mathbf{q} = \left[q_1,\cdots,q_K\right]^T$, $\mathbf{D} = \frac{1}{\tilde{g}_r} \times diag\left\{\gamma_1 / \left|\hat{h}_{1,1}\right|^2, \cdots, \gamma_K / \left|\hat{h}_{K,K}\right|^2\right\}$, $\mathbf{\Psi}$ is a coupling matrix with $\left[\mathbf{\Psi}\right]_{i,k} = \left|\hat{h}_{i,k}\right|^2$ if $i \neq k$, otherwise $\left[\mathbf{\Psi}\right]_{i,k} = 0$. $\mathbf{\sigma} = \left[\hat{\sigma}_1^2,\cdots,\hat{\sigma}_K^2\right]^T$ is a column vector stacking the $K$ noise variances. Combining (23) and the sum power constraint in (22), the solution of (22) can be solved by the following eigensystem

$$\mathbf{\Lambda}\mathbf{q}_{ext} = \frac{1}{C^U}\mathbf{q}_{ext}$$

(24)

where $\mathbf{q}_{ext} = \left[\mathbf{q}^T \quad 1\right]^T$ is the extended power vector, $\mathbf{\Lambda} = \begin{bmatrix} \mathbf{D\Psi}^T & \mathbf{D\sigma} \\ \frac{1}{P_b^{\max}}\mathbf{1}^T\mathbf{D\Psi}^T & \frac{1}{P_b^{\max}}\mathbf{1}^T\mathbf{D\sigma} \end{bmatrix}$ is an extended coupling matrix, with $\mathbf{1}$ being the all-one vector. From the Perron-Frobenius theory [16], we know that the optimal power vector $\mathbf{q}_{ext}$ is the unique positive eigenvector corresponding to the maximal eigenvalue of the nonnegative matrix $\mathbf{\Lambda}$, which is also the reciprocal of the balanced level $C^U$. Finally, $\mathbf{q}$ can be derived after scaling the vector $\mathbf{q}_{ext}$ such that the last component of $\mathbf{q}_{ext}$ equals to one. With (17), (21) and (24), an two-loop iterative algorithm that optimizes the beamformers and the uplink power allocation iteratively in the inner loop and the downlink power allocation in the outer loop is proposed. The procedures of the proposed algorithm are summarized in Table I(a) where the condition $t \leq 1$ means that the target SINRs are feasible.

TABLE I

| AF Based Relaying Scheme |
| --- |

| (a). Feasibility Testing Problem | (b). Sum Power Minimization Problem |
|---|---|
| **Initialize** | **Initialize** |
| $\quad \mathbf{q} = \dfrac{P_b^{\max}}{K}[1,\dots,1]^T$. | $\quad \mathbf{q}$ , $g_r$ : results from feasibility test |
| $\quad g_r = 1$ | $\quad m \leftarrow 0$ |
| $\quad n \leftarrow 0$ | $\quad$ **repeat** |
| **repeat** | $\qquad m \leftarrow m+1$ |
| $\quad n \leftarrow n+1$ | $\qquad$ Solve (17) with normalized uplink |
| $\quad m \leftarrow 0$ | $\qquad$ SINR, $1 \le k \le K$ . |
| $\quad$ **repeat** | $\qquad \mathbf{w}_k = \mathbf{w}_k / \|\mathbf{w}_k\|$, $1 \le k \le K$. |
| $\qquad m \leftarrow m+1$ | $\qquad$ Solve (29) to obtain $\mathbf{q}$ |
| $\qquad$ Solve (17) with normalized | $\quad$ **until** |
| $\qquad$ uplink SINR, $1 \le k \le K$ . | $\quad \big| \text{sum}(\mathbf{q}(m)) - \text{sum}(\mathbf{q}(m-1)) \big| < \varepsilon$ |
| $\qquad \mathbf{w}_k = \mathbf{w}_k / \|\mathbf{w}_k\|$, $1 \le k \le K$. | Solve (26) to obtain $t$ , $g_r$ and $\mathbf{p}$ |
| $\qquad$ Solve (24) to obtain $C^U$ and | |
| $\qquad \mathbf{q}_{ext}$ | |
| $\quad$ **until** | |
| $\qquad \big| C^U(m) - C^U(m-1) \big| < \varepsilon$ | |
| $\quad$ Solve (21) to obtain $t$ , $g_r$ and $\mathbf{p}$ | |
| **until** | |
| $\quad t(m) \le 1$ or $\big| t(m) - t(m-1) \big| < \varepsilon$ | |

*3) Power Allocation at the BS and the RS for Power Minimization Problem*

As soon as the feasibility test passes (i.e., $t \le 1$), the total power minimization problem can be carried out.

*a) Downlink Power Allocation*

By constraining $SINR_k = \gamma_k$, $k = 1,\dots,K$ , and with maximum power limitations at the BS and the RS, the power allocation problem which minimizes the total transmission power for a given $\mathbf{W}$ is

$$\min_{p_1,\dots,p_K,\,g_r} \quad \sum_{k=1}^{K} p_k + \sum_{k=1}^{K}\sum_{j=1}^{K} g_r p_k \left|\mathbf{H}_r(j,k)\right|^2 + K g_r \sigma_r^2$$

$$s.t. \quad \frac{p_k g_r \left|\hat{h}_{k,k}\right|^2}{\sum_{i=1,\neq k}^{K} p_i g_r \left|\hat{h}_{k,i}\right|^2 + g_r \sigma_r^2 \left\|\mathbf{g}_k^H\right\|^2 + \sigma_k^2} \ge \gamma_k, \ k = 1,\dots,K$$

$$\sum_{k=1}^{K} p_k \le P_b^{\max}$$

$$\sum_{k=1}^{K}\sum_{j=1}^{K} g_r p_k \left|\mathbf{H}_r(j,k)\right|^2 + K g_r \sigma_r^2 \le P_r^{\max}$$

(25)

The above optimization problem can also be converted into a GP problem by: 1.) Substituting the objective function with an extra variable $t$ to form a monomial objective function, and adding an additional posynomial inequality to the constraint set. 2.) Taking the inverse of the first inequality in (25) such that every equation in the constraint set is a posynomial. The standard GP formulation of this problem is

$$\min_{p_1,\ldots,p_k,g_r,t} \quad t$$

$$s.t. \quad \sum_{k=1}^{K} p_k + \sum_{k=1}^{K}\sum_{j=1}^{K} g_r p_k \left| \mathbf{H}_r(j,k) \right|^2 + Kg_r\sigma_r^2 \leq t$$

$$\frac{\sum_{i=1,\neq k}^{K} p_i g_r \left| \hat{h}_{k,i} \right|^2 + g_r\sigma_r^2 \left\| \mathbf{g}_k^H \right\|^2 + \sigma_k^2}{p_k g_r \left| \hat{h}_{k,k} \right|^2} \leq \frac{1}{\gamma_k}, \; k=1,\ldots,K$$

$$\sum_{k=1}^{K} p_k \leq P_b^{\max}$$

$$\sum_{k=1}^{K}\sum_{j=1}^{K} g_r p_k \left| \mathbf{H}_r(j,k) \right|^2 + Kg_r\sigma_r^2 \leq P_r^{\max}$$

$$(26)$$

*b)    Uplink Power Allocation*

For the virtual uplink power minimization problem with the SINR targets $\lambda_k$, $k=1,\ldots,K$, and given beamformer $\tilde{\mathbf{W}}$ and $\tilde{g}_r$, we have the following problem formulation for obtaining the optimal power allocation at MSs.

$$\min_{q_1,\ldots,q_K} \quad \sum_{k=1}^{K} q_k$$

$$s.t. \quad \frac{q_k \tilde{g}_r \left| \hat{h}_{k,k} \right|^2}{\left( \sum_{i=1,\neq k}^{K} q_i \tilde{g}_r \left| \hat{h}_{i,k} \right|^2 + \hat{\sigma}_k^2 \right)} \geq \gamma_k$$

$$(27)$$

Due to the monotonic properties of $SINR_k$ discussed before, the inequality constraint in (27) must be met with equality, and the power allocation vector $\mathbf{q}$ that fulfills this design goal can be obtained by solving the following equation [12]

$$\mathbf{q} = \mathbf{D}\mathbf{\Psi}^T \mathbf{q} + \mathbf{D}\boldsymbol{\sigma} , \tag{28}$$

with the same parameters defined in (23). The solution can be derived as

$$\mathbf{q} = \left( \mathbf{I} - \mathbf{D}\mathbf{\Psi}^T \right)^{-1} \mathbf{D}\boldsymbol{\sigma} . \tag{29}$$

Note that this solution is derived based on the same total power consumed at the BS and the same beamformers $\tilde{\mathbf{W}}$ and $\tilde{g}_r$ as in the downlink power allocation. It is known that a feasible positive solution of $\mathbf{q}$ exists if the spectral radius of $\mathbf{D}\mathbf{\Psi}^T$ is smaller than one, i.e., $\lambda_{\max}(\mathbf{D}\mathbf{\Psi}^T) < 1$ [16], where $\lambda_{\max}(\mathbf{M})$ denotes the maximum eigenvalue of matrix $\mathbf{M}$. This can be ensured by the feasibility test discussed before. Once the feasibility test passes, the algorithm can switch to the sum power minimization problem. The procedures for the sum power minimization problem are given in Table I(b). It can be seen that there is no outer-loop iteration defined in Table I(b). The reason is that the value of the RS gain factor $g_r$ in (26) has no effect on the beamformer weights $\mathbf{w}_k$, $1 \leq k \leq K$ provided that $g_r \neq 0$. The influence of $g_r$ in the normalized virtual uplink channel is a real scalar

$$\frac{g_r}{\hat{\sigma}_k} = \frac{g_r}{\sqrt{\sigma_r^2 g_r \left\| \mathbf{g}_k^H \right\|^2 + \sigma_k^2}}$$ which is multiplied by the equivalent channel vector $\mathbf{H}^H \mathbf{g}_k$, therefore, the spatial feature of the channel

vector is unchanged. Besides, the scaling multiplier will be incorporate into the uplink power allocation stage and its effect can be compensated via $g_r$-dependent power allocation to reach the same individual SINR constraint.

### B. SVD Based Relaying with Downlink-Uplink Duality at the RS

In this subsection, we consider another relaying strategy that operates on the eigenspaces of the MIMO channel between the BS and the RS. The numbers of antennas at the BS and the RS are assumed to be larger than the number of users. In this scheme, the RS conducts more complex signal processing rather than the simple AF, and it should be noted that the downlink-uplink duality here is established for the RS-MS channel. With singular value decomposition (SVD) of the channel matrix $\mathbf{H}$, we have

$$\mathbf{H} = \mathbf{U}\boldsymbol{\Sigma}\mathbf{V}^H \tag{30}$$

where $\mathbf{U}$ and $\mathbf{V}$ are unitary matrices, and $\boldsymbol{\Sigma}$ is a diagonal matrix with singular values which are all nonnegative. Without loss of generality, let the nonzero elements of $\boldsymbol{\Sigma}$ be in descending order, i.e. $\lambda_1 \geq \lambda_2 \geq ... \geq \lambda_{\min(M_b, M_r)}$. We can collect the eigenspaces corresponding to the largest $K$ singular values as our $K$ spatial parallel sub-channels. For notational simplicity, we assume the case of full-rank MIMO channel with $M_b = M_r = K$ in the following derivations. The precoding matrix at the BS is constructed using the unitary matrix $\mathbf{V}$, such that

$$\mathbf{F}\mathbf{F}^H = \mathbf{V}\boldsymbol{\Lambda}_p\mathbf{V}^H \tag{31}$$

where $\boldsymbol{\Lambda}_p = diag\left\{ p_1, ..., p_K \right\}$. The linear processing matrix $\mathbf{Q}$ at the RS is configured as

$$\mathbf{Q} = \mathbf{A}\boldsymbol{\Lambda}_r^{1/2}\mathbf{U}^H \tag{32}$$

where the matrix $\mathbf{U}$ is for receive beamforming which is derived from the SVD of $\mathbf{H}$. The matrix $\mathbf{A}$ represents the transmit beamforming matrix consisting of unit norm transmit beamformers, i.e., $\mathbf{A} = [\mathbf{a}_1, ..., \mathbf{a}_K]$ which will be derived later, and the diagonal matrix $\boldsymbol{\Lambda}_r$ consists of elements of the power allocation vector $\mathbf{p}^r = \left[ p_1^r, ..., p_K^r \right]$ at the RS for the $K$ streams. Note that this SVD structure divides the BS-RS MIMO channel into parallel subchannels, and allows the RS to further perform pairing of its incoming and outgoing subchannels. In the later part of this section, a subchannel pairing scheme which enhances power efficiency will be discussed.

The $k$th stream after receive beamforming at the RS can be expressed as

$$r_k = \lambda_k \sqrt{p_k} s_k + \mathbf{u}_k^H \mathbf{n}_r \tag{33}$$

where $\mathbf{U} = [\mathbf{u}_1,...,\mathbf{u}_K]$. Since $\mathbf{U}$ is a unitary matrix, the noise power after receive beamforming at RS remains $\sigma_r^2$ for each stream. To normalize the received power at the RS, a normalization scaling factor

$$\varepsilon_k = \sqrt{1/\left(\lambda_k^2 p_k + \sigma_r^2\right)} \tag{34}$$

is introduced such that the $k$th received stream after normalization equals to

$$\hat{r}_k = \varepsilon_k r_k \tag{35}$$

which has unit signal power, i.e., $|\hat{r}_k|^2 = 1$. Therefore, the power consumption at the RS can be simplified to

$$\mathrm{Tr}\left\{\mathbf{A}\left(\mathbf{\Lambda}_r^{1/2}\hat{\mathbf{R}}\mathbf{\Lambda}_r^{1/2}\right)\mathbf{A}^H\right\} = \mathrm{Tr}\left\{\mathbf{\Lambda}_r\right\} = \sum_{k=1}^{K} p_k^r \tag{36}$$

where $\hat{\mathbf{R}} = diag\left\{\hat{r}_1,...,\hat{r}_K\right\}$. The received stream at the $k$th MS can be expressed as

$$y_k = \mathbf{g}_k^H \mathbf{a}_k \sqrt{p_k^r} \hat{r}_k + \mathbf{g}_k^H \sum_{i=1,\neq k}^{K} \mathbf{a}_i \sqrt{p_i^r} \hat{r}_i + n_k. \tag{37}$$

Based on (33) to (35) and (37), we have the following SINR expression for the $k$th stream

$$SINR_k = \frac{\left|\mathbf{g}_k^H \mathbf{a}_k\right|^2 p_k^r \varepsilon_k^2 \lambda_k^2 p_k}{\sum_{i=1,\neq k}^{K} \left|\mathbf{g}_k^H \mathbf{a}_i\right|^2 p_i^r \varepsilon_i^2 \lambda_i^2 p_i + \sum_{j=1}^{K} \left|\mathbf{g}_k^H \mathbf{a}_j\right|^2 p_j^r \varepsilon_j^2 \sigma_r^2 + \sigma_k^2}. \tag{38}$$

Again, the optimization problems in (10) and (11) with the above SINR results are not jointly convex cases for the power allocation parameters $\mathbf{\Lambda}_p, \mathbf{\Lambda}_r$ and the transmit beamformer matrix $\mathbf{A}$. In the following, a design criterion based on the downlink-uplink duality at the RS is proposed such that the original problem can be divided into the power allocation problem (for $\mathbf{\Lambda}_p, \mathbf{\Lambda}_r$) and the transmit beamformer design problem at the RS (for $\mathbf{A}$).

### 1) Beamfomer Design via Downlink-Uplink Duality at the RS

In this section, we formulate an optimization problem with respect to the transmit beamformer $\mathbf{A}$ at the RS by assuming that the power allocations at the BS and the RS are fixed. The achieved SINR of the $k$th stream after receive beamforming at the relay can be expressed as

$$\alpha_k = \frac{p_k \lambda_k^2}{\sigma_r^2}. \tag{39}$$

With given power allocation matrices $\mathbf{\Lambda}_p$ and $\mathbf{\Lambda}_r$, i.e., $p_k$ is fixed to $\tilde{p}_k$ and $p_k^r$ is fixed to $\tilde{p}_k^r$ for $k \in \{1,\cdots,K\}$, we have $\tilde{\alpha}_k = \frac{\tilde{p}_k \lambda_k^2}{\sigma_r^2}$, and the downlink SINR at the $k$th MS in (38) can be rewritten as

$$SINR_k^D = \frac{\dfrac{\tilde{\alpha}_k}{1+\tilde{\alpha}_k}\,\tilde{p}_k^r \mathbf{g}_k^H \mathbf{a}_k \mathbf{a}_k^H \mathbf{g}_k}{\mathbf{g}_k^H \left( \displaystyle\sum_{i=1,\neq k}^{K} \tilde{p}_i^r \mathbf{a}_i \mathbf{a}_i^H + \dfrac{1}{1+\tilde{\alpha}_k}\,\tilde{p}_k^r \mathbf{a}_k \mathbf{a}_k^H \right)\mathbf{g}_k + \sigma_k^2} \ . \tag{40}$$

The first term of the denominator is the sum of interferences caused by the coupling at the second hop, and the second term of the denominator is the consequence of the amplified noise at the RS. The coupling vectors $\mathbf{a}_i, i \in \{1,...,K\}\setminus\{k\}$ in the denominator makes direct computing of all vectors difficult. However, based on the concept of downlink-uplink duality, we can solve the problem using the equivalent virtual uplink channel. The SINR expression of the virtual uplink channel for the $k$th MS is

$$SINR_k^U = \frac{\mathbf{a}_k^H \left( \dfrac{\tilde{\alpha}_k}{1+\tilde{\alpha}_k}\,\tilde{q}_k^r \mathbf{g}_k \mathbf{g}_k^H \right)\mathbf{a}_k}{\mathbf{a}_k^H \left( \displaystyle\sum_{i=1,\neq k}^{K} \tilde{q}_i^r \mathbf{g}_i \mathbf{g}_i^H + \dfrac{1}{1+\tilde{\alpha}_k}\,\tilde{q}_k^r \mathbf{g}_k \mathbf{g}_k^H + \sigma_k^2 \right)\mathbf{a}_k} \tag{41}$$

where $q_k^r$, $k = 1,...,K$, denotes the corresponding power allocation in the virtual uplink channel. Using the virtual uplink, the beamformers $\mathbf{a}_k$, $k \in \{1,\cdots,K\}$ can be derived by solving $K$ decoupled generalized eigenvalue problems, i.e.,

$$\mathbf{a}_k^{opt} = \arg\max_{\lambda} Eig\left( \mathbf{R}_{s_k}, \mathbf{R}_{n_k} \right) \tag{42}$$

where $\mathbf{R}_{s_k} = \dfrac{\tilde{\alpha}_k}{1+\tilde{\alpha}_k}\,\tilde{q}_k^r \mathbf{g}_k \mathbf{g}_k^H$ and $\mathbf{R}_{n_k} = \displaystyle\sum_{i=1,\neq k}^{K} \tilde{q}_i^r \mathbf{g}_i \mathbf{g}_i^H + \dfrac{1}{1+\tilde{\alpha}_k}\,\tilde{q}_k^r \mathbf{g}_k \mathbf{g}_k^H + \sigma_k^2$ are the desired signal covariance matrix and the undesired signal covariance matrix of the $k$th stream, respectively.

Compared to the MIMO broadcast channel without relay, in this structure, in addition to the coupling effect, the achieved SINR $\alpha_k$, $k = 1,...,K$ at the first hop and the amplified noise at the second hop are also taken into account in our derivation.

### 2) Power Allocation for Feasibility Test Problem

#### a) Downlink Power Allocation

Assuming fixed transmit beamforming matrix at the RS, i.e., $\mathbf{A} = \tilde{\mathbf{A}} = [\tilde{\mathbf{a}}_1,...,\tilde{\mathbf{a}}_K]$ with $\|\tilde{\mathbf{a}}_k\| = 1$ for all $k = 1,...,K$, the downlink SINR of the $k$th MS in (38) can be expressed as

$$SINR_k = \frac{\left|\hat{g}_{k,k}\right|^2 p_k^r \varepsilon_k \lambda_k^2 p_k}{\displaystyle\sum_{i=1,\neq k}^{K} \left|\hat{g}_{k,i}\right|^2 p_i^r \varepsilon_i \lambda_i^2 p_i + \sum_{j=1}^{K} \left|\hat{g}_{k,j}\right|^2 p_j^r \varepsilon_j \sigma_r^2 + \sigma_k^2} \tag{43}$$

where $\hat{g}_{k,i} = \mathbf{g}_k^H \tilde{\mathbf{a}}_i$. From (34), (43) and with some manipulations, the optimization problem for feasibility test in (11) can be reformulated as

$$\max_{\substack{p_1,\ldots,p_K \\ p_1^r,\ldots,p_K^r}} \min_{k \in \{1,\ldots,K\}}$$

$$\frac{\left|\hat{g}_{k,k}\right|^2 \lambda_k^2 p_k p_k^r}{\gamma_k \left\{\left(\lambda_k^2 p_k + \sigma_r^2\right)\left(\sum_{i=1,\neq k}^K p_i^r \left|\hat{g}_{k,i}\right|^2 + \sigma_k^2\right) + p_k^r \left|\hat{g}_{k,k}\right|^2 \sigma_r^2\right\}} . \tag{44}$$

$$subject\ to$$

$$\sum_{k=1}^K p_k \leq P_b^{\max} \quad , \sum_{k=1}^K p_k^r \leq P_r^{\max}$$

By introducing a new design variable $t$, we can equivalently rewrite the above optimization problem as follows

$$\min_{p_k, p_k^r, \forall k, t} \quad t$$

$$subject\ to$$

$$\frac{\gamma_k \left\{\left(\lambda_k^2 p_k + \sigma_r^2\right)\left(\sum_{i=1,\neq k}^K p_i^r \left|\hat{g}_{k,i}\right|^2 + \sigma_k^2\right) + p_k^r \left|\hat{g}_{k,k}\right|^2 \sigma_r^2\right\}}{\left|\hat{g}_{k,k}\right|^2 \lambda_k^2 p_k p_k^r} \leq t,$$

$$k = 1,\ldots,K,$$

$$\sum_{k=1}^K p_k \leq P_b^{\max} \quad , \sum_{k=1}^K p_k^r \leq P_r^{\max}. \tag{45}$$

Since the objective function in (45) is a monomial, and all the constraints are posynomial inequalities, the above optimization problem is a standard form of GP problem.

*b)    Uplink Power Allocation*

The power allocation in the virtual uplink channel for the SINR balancing problem can be easily formulated as

$$\max_{q_1^r,\ldots,q_K^r} \min_{k \in \{1,\ldots,K\}} \frac{SINR_k^U}{\gamma_k}$$

$$subject\ to\ \sum_{k=1}^K q_k^r \leq P_r^{\max} \tag{46}$$

where only the relay power constraint is considered here, and the transmission power at the BS and the corresponding constraint are obtained in the downlink power allocation step. The expression of uplink SINR can be derived based on (41)

$$SINR_k^U = \frac{\frac{\tilde{\alpha}_k}{1+\tilde{\alpha}_k} q_k^r \left|\hat{\mathbf{g}}_{k,k}\right|^2}{\sum_{i=1}^K q_i^r \left|\hat{\mathbf{g}}_{k,k}\right|^2 + \frac{1}{1+\tilde{\alpha}_k} q_k^r \left|\hat{\mathbf{g}}_{k,k}\right|^2 + \sigma_k^2} . \tag{47}$$

Since the same achieved SINR balanced level of all MSs must be met with equality of the constraint in (46) due to the monotonic properties of $SINR_k^U$, with the definition of a common balance level $C^U = \frac{SINR_k^U}{\gamma_k}$ for MS $k$, the optimal uplink power vector can be derived by substituting (47) to the objective function of (46), with equality of the constraint in (46). The result is an eigensystem

$$\mathbf{\Lambda q}_{ext}^r = \frac{1}{C^U}\mathbf{q}_{ext}^r \qquad (48)$$

where $\mathbf{q}_{ext}^r = \left[\left(\mathbf{q}^r\right)^T \quad 1\right]^T$ is the extended power vector, and $\mathbf{\Lambda}$ is the extended coupling matrix expressed as

$$\mathbf{\Lambda} = \begin{bmatrix} \mathbf{D}\mathbf{\Psi}^T + \mathbf{E} & \mathbf{D}\boldsymbol{\sigma} \\ \dfrac{1}{P_r^{\max}}\mathbf{1}^T\left(\mathbf{D}\mathbf{\Psi}^T + \mathbf{E}\right) & \dfrac{1}{P_r^{\max}}\mathbf{1}^T\mathbf{D}\boldsymbol{\sigma} \end{bmatrix} \qquad (49)$$

where $\mathbf{D} = diag\left\{\dfrac{\gamma_1\left(1+\tilde{\alpha}_1\right)}{\tilde{\alpha}_1\left|\hat{\mathbf{g}}_{1,1}\right|^2},\cdots,\dfrac{\gamma_K\left(1+\tilde{\alpha}_K\right)}{\tilde{\alpha}_K\left|\hat{\mathbf{g}}_{K,K}\right|^2}\right\}$, and $\mathbf{E} = diag\left\{\gamma_1/\tilde{\alpha}_1,\cdots,\gamma_K/\tilde{\alpha}_K\right\}$, $\left[\mathbf{\Psi}\right]_{ij} = \left|\hat{\mathbf{g}}_{i,j}\right|^2$ if $j \neq i$, otherwise $= 0$, $\boldsymbol{\sigma} = \sigma_d^2\mathbf{1}$ is

the $K\times1$ vector of the destination noise power and $\mathbf{1} = [1,\cdots,1]^T$. The optimal power vector $\mathbf{q}_{ext}^r$ is a unique positive eigenvector corresponding to the maximal eigenvalue of the nonnegative matrix $\mathbf{\Lambda}$ [16], which is also the reciprocal of the balanced level $C^U$.

The optimal power vector $\mathbf{q}^r$ is then obtained by scaling $\mathbf{q}_{ext}^r$ such that the last component of $\mathbf{q}_{ext}^r$ equals to one. It can be observed from (49) that, when $\tilde{\alpha}_k \to \infty$ for $k = 1,\ldots,K$, that is, there is no noise term induced from the first hop, the results of our derivations reduce to the framework proposed in [12], which can be thought of as a special condition of our system. Based on (42), (45) and (48), an iterative updating algorithm that iterates between the beamformer, the uplink power allocation, and the downlink power allocation optimizations is proposed. The procedures are summarized in Table II(a).

TALBE II

| SVD Based Relaying Scheme | |
| --- | --- |
| (a). Feasibility Testing Problem | (b). Sum Power Minimization Problem |
| **Initialize**<br>$\mathbf{p} = \dfrac{P_b^{\max}}{K}[1,\ldots,1]^T$.<br><br>$\mathbf{q}^r = \dfrac{P_r^{\max}}{K}[1,\ldots,1]^T$<br><br>$n \leftarrow 0$<br>**repeat**<br>$\quad n \leftarrow n+1$<br>$\quad m \leftarrow 0$<br>$\quad$**repeat**<br>$\quad\quad m \leftarrow m+1$<br>$\quad\quad$Calculate (39) to obtain<br>$\quad\quad\alpha_k,\ 1 \leq k \leq K$<br>$\quad\quad$Solve (42) with equivalent<br>$\quad\quad$uplink SINR, $1 \leq k \leq K$.<br>$\quad\quad\mathbf{a}_k = \mathbf{a}_k/\|\mathbf{a}_k\|,\ 1 \leq k \leq K$.<br>$\quad\quad$Solve (48) to obtain<br>$\quad\quad C^U$ and $\mathbf{q}_{ext}^r$<br>$\quad$**until** | **Initialize**<br>$\mathbf{p},\mathbf{q}^r$: results form feasibility test<br>$n \leftarrow 0$<br>**repeat**<br>$\quad n \leftarrow n+1$<br>$\quad m \leftarrow 0$<br>$\quad$**repeat**<br>$\quad\quad m \leftarrow m+1$<br>$\quad\quad$Calculate (39) to obtain<br>$\quad\quad\alpha_k,\ 1 \leq k \leq K$<br>$\quad\quad$Solve (42) with equivalent<br>$\quad\quad$uplink SINR, $1 \leq k \leq K$.<br>$\quad\quad\mathbf{a}_k = \mathbf{a}_k/\|\mathbf{a}_k\|,\ 1 \leq k \leq K$.<br>$\quad\quad$Solve (53) to obtain $\mathbf{q}^r$<br>$\quad$**until**<br>$\quad\left|\text{sum}(\mathbf{q}^r(m)) - \text{sum}(\mathbf{q}^r(m-1))\right| < \varepsilon$<br>Solve (51) to obtain $t$, $\mathbf{p}$ and $\mathbf{p}^r$<br>**until** |

| $\left\|C^U(m)-C^U(m-1)\right\|<\varepsilon$ | $\left\|t(m)-t(m-1)\right\|<\varepsilon$ |
|---|---|
| Solve (45) to obtain $t$, $\mathbf{p}$ and $\mathbf{p}^r$ **until** $t(m)\leq 1$ or $\left\|t(m)-t(m-1)\right\|<\varepsilon$ | |

### 3) Power Allocation for the Power Minimization Problem

If the feasibility test passes, the algorithm can switch to the sum power minimization problem.

#### a)  Downlink Power Allocation

Based on the result in (43), the power minimization problem of MIMO relay downlink channel can be expressed as

$$\min_{\mathbf{p},\mathbf{p}^r}\quad \sum_{k=1}^{K}p_k+\sum_{k=1}^{K}p_k^r$$
$$s.t.$$
$$\frac{\left|\hat{g}_{k,k}\right|^2\lambda_k^2 p_k p_k^r}{\gamma_k\left\{\left(\lambda_k^2 p_k+\sigma_r^2\right)\left(\sum_{i=1,\neq k}^{K}p_i^r\left|\hat{g}_{k,i}\right|^2+\sigma_k^2\right)+p_k^r\left|\hat{g}_{k,k}\right|^2\sigma_r^2\right\}}\geq 1,$$
$$k=1,\ldots,K,$$
$$\sum_{k=1}^{K}p_k\leq P_b^{\max},\sum_{k=1}^{K}p_k^r\leq P_r^{\max}$$

(50)

It can be observed that the above problem can also be recast as a GP problem after introducing another variable $t$, similar to that in the case of the simple AF based relaying design. The standard GP formulation of this problem is as follows

$$\min_{\mathbf{p},\mathbf{p}^r}\quad t$$
$$s.t.\ \sum_{k=1}^{K}p_k+\sum_{k=1}^{K}p_k^r\leq t$$
$$\frac{\left(\lambda_k^2 p_k+\sigma_r^2\right)\left(\sum_{i=1,\neq k}^{K}p_i^r\left|\hat{g}_{k,i}\right|^2+\sigma_k^2\right)+p_k^r\left|\hat{g}_{k,k}\right|^2\sigma_r^2}{\left|\hat{g}_{k,k}\right|^2\lambda_k^2 p_k p_k^r}\leq\gamma_k,$$
$$k=1,\ldots,K$$
$$\sum_{k=1}^{K}p_k\leq P_b^{\max}\ ,\sum_{k=1}^{K}p_k^r\leq P_r^{\max}.$$

(51)

#### b)  Uplink  Power Allocation

The virtual uplink power allocation for the power minimization problem can be formulated  as

$$\min\sum_{k=1}^{K}q_k^r$$
$$subject\ to\ \frac{SINR_k^U}{\gamma_k}\geq 1,\ k=1,\ldots,K.$$

(52)

By substituting (47) into the power minimization problem (52), due to the monotonic properties of $SINR_k^{U}$, the constraint in (52) must be met with equality. The optimal solution of $\mathbf{q}^r = \left[ q_1^r, \ldots, q_k^r \right]^T$ can then be expressed as

$$\mathbf{q}^r = \left[ \mathbf{I} - (\mathbf{D}\boldsymbol{\Psi}^T + \mathbf{E}) \right]^{-1} \mathbf{D}\boldsymbol{\sigma} \tag{53}$$

with the same parameters defined in (49). By examining (53) and defining $\mathbf{Z} = \mathbf{D}\boldsymbol{\Psi}^T + \mathbf{E}$, the necessary and sufficient condition to have positive solution $\mathbf{q}^r$ for any positive vector $\mathbf{D}\boldsymbol{\sigma}$ is that $(\mathbf{I} - \mathbf{Z})^{-1}$ must be nonnegative. Since the matrix $\mathbf{Z}$ is positive definite, in order to realize $(\mathbf{I} - \mathbf{Z})^{-1} \geq \mathbf{0}$, the spectral radius of $\mathbf{Z}$ must be smaller than one, i.e., $\rho(\mathbf{Z}) = \lambda_{\max}(\mathbf{Z}) < 1$. With this constraint, $(\mathbf{I} - \mathbf{Z})^{-1} = \sum_{k=0}^{\infty} \mathbf{Z}^k$ will converge to a finite nonnegative matrix. From Gersgorin's theory [16], we known that $\left| \lambda_i - z_{ii} \right| \leq \sum_{i \neq j} \left| z_{ij} \right|$, where $\lambda_i$ is the $i$th eigenvalue of the corresponding eigenvector of $\mathbf{Z}$. Substituting $\mathbf{Z}$ with $\mathbf{D}\boldsymbol{\Psi}^T + \mathbf{E}$, we have

$$\left| \lambda_k - \frac{\gamma_k}{\tilde{\alpha}_k} \right| \leq \frac{\gamma_k (1 + \tilde{\alpha}_k)}{\tilde{\alpha}_k} \sum_{i \neq k} \frac{\left| \hat{g}_{i,k} \right|^2}{\left| \hat{g}_{k,k} \right|^2}. \tag{54}$$

By constraining the maximum possible value of $\lambda_k$ smaller than one in (54), we have the following result for $\gamma_k$

$$\gamma_k < \frac{\tilde{\alpha}_k \left| \hat{g}_{k,k} \right|^2}{\left| \hat{g}_{k,k} \right|^2 + (1 + \tilde{\alpha}_k) \sum_{i \neq k} \left| \hat{g}_{i,k} \right|^2}, \ k = 1, \ldots, K \tag{55}$$

which can be simplified to

$$\gamma_k < \frac{\tilde{\alpha}_k \chi_k}{1 + \tilde{\alpha}_k + \chi_k}, \ k = 1, \ldots, K \tag{56}$$

where

$$\chi_k = \sum_{i \neq k} \frac{\left| \hat{g}_{k,k} \right|^2}{\left| \hat{g}_{i,k} \right|^2}. \tag{57}$$

Thus, to have positive power allocation $\mathbf{q}^r$, $\alpha_k, \gamma_k$ and $\chi_k$ should satisfy the inequality (56), which depends on the achieved SINR in the first hop and the spatial separability in the second hop. It can be also observed that the values of noise variances received at the MSs have no effect on the feasibility condition of $\mathbf{q}^r$ in (56). In fact, the inequality constraint (55) represents only a sufficient condition to obtain a feasible solution, which may be too restrictive. With the help of the feasibility test problem, not only the positive results of $\mathbf{q}^r$ are ensured, unlimited power (when $\mathbf{D}\boldsymbol{\Psi}^T + \mathbf{E}$ is close to $\mathbf{I}$) in (53) is also prevented.

In order to prove that the solution of (53) achieves the same SINR region as the downlink channel under the same amount of power at RS, we now consider the downlink problem with given $\tilde{\alpha}_k$ as

$$\min \sum_{k=1}^{K} p_k^r$$
$$subject\ to\ \frac{SINR_k^D}{\gamma_k} \geq 1,\ k = 1,...,K \tag{58}$$

where

$$SINR_k^D = \frac{\frac{\tilde{\alpha}_k}{1+\tilde{\alpha}_k} p_k^r \left|\hat{\mathbf{g}}_{k,k}\right|^2}{\sum_{i=1}^{K} p_i^r \left|\hat{\mathbf{g}}_{k,i}\right|^2 + \frac{1}{1+\tilde{\alpha}_k} p_k^r \left|\hat{\mathbf{g}}_{k,k}\right|^2 + \sigma_k^2}\ . \tag{59}$$

Similar to (52), the optimum downlink power allocation vector at the RS has the following result

$$\mathbf{p}^r = \left[\mathbf{I} - (\mathbf{D\Psi} + \mathbf{E})\right]^{-1} \mathbf{D\sigma}. \tag{60}$$

By taking the $L_1$-norm operation on (53) and (60), we have

$$\left\|\mathbf{p}^r\right\|_1 = \mathbf{1}^T \mathbf{p}^r = \mathbf{1}^T (\mathbf{I} - \mathbf{D\Psi} - \mathbf{E})^{-1} \mathbf{D\sigma}$$
$$= \mathbf{1}^T (\mathbf{I} - \mathbf{D\Psi}^T - \mathbf{E})^{-1} \mathbf{D\sigma} = \left\|\mathbf{q}^r\right\|_1\ . \tag{61}$$

Since $\mathbf{E}$ is a diagonal matrix with positive real elements, we have $\lambda_{\max}\left(\mathbf{D\Psi} + \mathbf{E}\right) = \lambda_{\max}\left(\mathbf{D\Psi}^T + \mathbf{E}\right)$ [12]. Therefore, the same SINR regions can be achieved with the same sum power at the RS, i.e., $\sum_{k=1}^{K} p_k^r = \sum_{k=1}^{K} q_k^r$. An immediate consequence of the above result is that: in the SVD based relaying, with given power allocation at the BS and sum power constraints at the RS, RS based downlink and virtual uplink channels have the same achievable SINR region with the same set of transmit beamformers at RS. Although the equality in (61) hold with the assumption of equal received noise power at each MS, for different received noise levels, i.e., $\sigma_1 \neq \sigma_2,...,\neq \sigma_K$, we can use the concept of normalized downlink channel to solve the equivalent uplink problem and achieve the optimal result. The procedures of this sum power minimization problem are summarized in Table II(b).

### 4) Subchannel Pairing

Up to now we have assumed that the $k$th signal stream transmitted over the $k$th subchannel (eigenspace) in the first hop is retransmitted by the RS on the $k$th subchannel in the second hop to the $k$th MS. Better power efficiency can be achieved if the subchannels of the two hops are paired according to their respective channel conditions. The authors in [11] adopted this idea under the scenario of joint ZF at both hops. The idea there was that the penalty of bad channel in either hop can be compensated by a good channel in the other hop to make the power allocation more efficient. In the following, we validate the performance gain of this idea for the proposed SVD based relaying scheme using a simple two-user case. From (39), the SINR expression of the $k$th MS in (38) can be rewritten as

$$SINR_k = \frac{\alpha_k \beta_k}{1 + \alpha_k + \beta_k}, \; k = 1,...,K \tag{62}$$

where

$$\beta_k = \frac{p_k^r \left| \hat{g}_{k,k} \right|^2}{\sum_{i=1, \neq k}^{K} p_i^r \left| \hat{g}_{k,i} \right|^2 + \sigma_k^2} \tag{63}$$

is the achieved SINR at the second hop. We define a channel to interference and noise ratio (CINR), which is the received SINR before power allocation. For the two-user case, the CINRs of the first hop subchannels are $CINR_k^1 = \lambda_k / \sigma_r^2$ for $k= 1$ and 2, and the

CINRs of the second hop subchannels are $CINR_k^2 = \sum_{i \neq k} \frac{\left| \hat{g}_{1,1} \right|^2}{\left| \hat{g}_{i,k} \right|^2 + \sigma_k^2}$ for $k = 1$ and 2. For ease of illustration, if we ignore the coupling

effect at the second hop, the behavior of subchannel pairing can be analyzed through the following power allocation problem, which is also a GP problem.

$$\min_{p_1, p_2, q_1, q_2} \quad p_1 + p_2 + q_1 + q_2$$
$$s.t.$$
$$\frac{p_1 q_1 CINR_1^1 CINR_1^2}{1 + p_1 CINR_1^1 + q_1 CINR_1^2} \geq \gamma_1, \tag{64}$$
$$\frac{p_2 q_2 CINR_2^1 CINR_2^2}{1 + p_2 CINR_2^1 + q_2 CINR_2^2} \geq \gamma_2$$

The minimum power levels that is required to achieve target $\gamma_1 = \gamma_2 = 1$ in problem (64) for various settings of $CINR_1^1$ and $CINR_2^1$ is shown in Fig. 2(a). Five discrete values of $CINR_1^1$ and $CINR_2^1$ $(0dB, 5dB, 10dB, 15dB, 20dB)$ are chosen for analysis. The result of solid line is under the assumption that $\left( CINR_1^2, CINR_2^2 \right) = (20dB, 0dB)$ while the dotted line is under the assumption $\left( CINR_1^2, CINR_2^2 \right) = (10dB, 5dB)$, both based on the assumption that $CINR_1^2 > CINR_2^2$. It can be observed that less total power is consumed for all the conditions with $CINR_1^1 < CINR_2^1$. This is consistent with the concept of channel compensation or the observations in [11]. That is, a better channel condition in the first hop should be paired with a worse channel condition in the second hop and vice versa. Although this rule may not be always correct if users have different target values, i.e., $\gamma_1 \neq \gamma_2$ (Fig. 2(b) shows the exceptional case for $\gamma_1 = 1$ and $\gamma_2 = 0.1$), in general, for $K > 2$, a heuristic subchannel pairing scheme is such that $CINR_{(1)}^1 \geq CINR_{(2)}^1 \geq \cdots \geq CINR_{(K)}^1$ is paired with $CINR_{(K)}^2 \leq CINR_{(K-1)}^2 \leq \cdots \leq CINR_{(1)}^2$, where the subscript (in increasing order) denotes the CINR in non-increasing order.

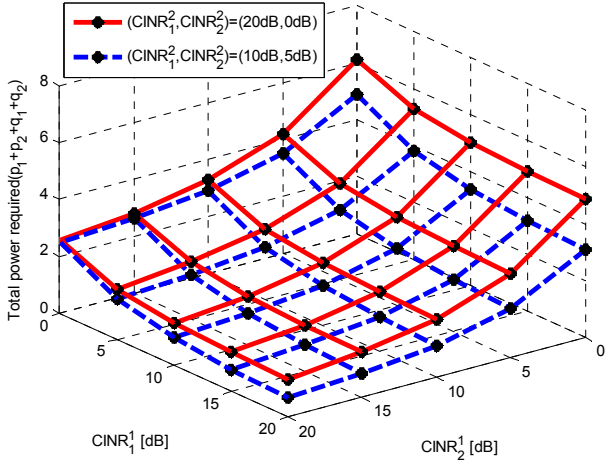

Fig. 2(a). Analysis of subchannel pairing for $\gamma_1 = \gamma_2 = 1$.

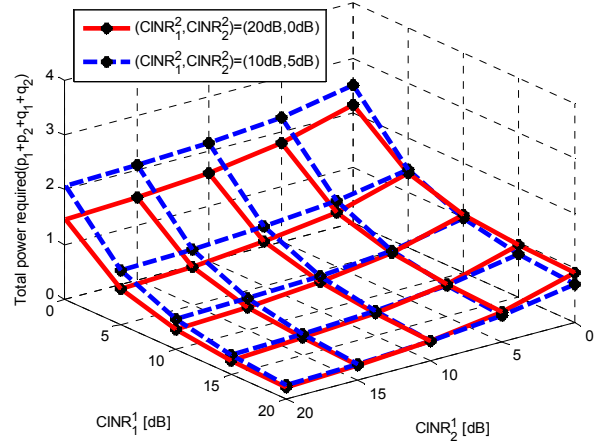

Fig. 2(b). Analysis of subchannel pairing for $\gamma_1 = 1$ and $\gamma_2 = 0.1$.

### 5) Generalization to Multi-hop MIMO Relays

The proposed scheme can also be extended to a multi-hop MIMO scenario with multiple cascaded MIMO channels as shown in Fig.3.

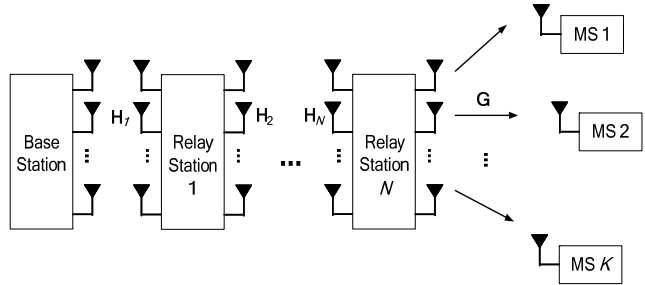

Fig. 3. Multihop MIMO relay downlink channel.

Multihop transmissions not only extend the coverage area but also improve the power efficiency [17]. If the total number of RSs is $N$, then there will be totally $N+1$ hops (with $N$ hops point-to-point MIMO channels and one-hop MIMO broadcast channel). The SVD decomposition is implemented in each point-to-point MIMO channel, and the optimum transmit beamformer design is conducted at the $N$th RS using the result derived in Section IV.B. If we define the equivalent SINR from the first hop to the $(n-1)$th hop as $SINR_{1 \sim (n-1)}$ and the SINR at the $n$th hop as $SINR_n$, the equivalent SINR from the first hop to the $n$th hop ($SINR_{1 \sim n}$) for non-regenerative relay systems with receive power normalization can be expressed as the following recursive equation [18]

$$SINR_{1\sim n} = \frac{SINR_{1\sim(n-1)}SINR_n}{1 + SINR_{1\sim(n-1)} + SINR_n} \quad . \tag{65}$$

With the above recursive property of the equivalent SINR, we have

$$SINR_{1\sim n}^{-1} = SINR_{1\sim(n-1)}^{-1}SINR_n^{-1} + SINR_{1\sim(n-1)}^{-1} + SINR_n^{-1}. \tag{66}$$

From (66), it can be observed that if $SINR_{1\sim(n-1)}^{-1}$ and $SINR_n^{-1}$ are both posynomials, then $SINR_{1\sim n}^{-1}$ is also a posynomial. Therefore, from (62) and (63), if we substitute the inverse of the first hop SINR $\alpha_k^{-1}$ in the two-hop system in Section IV.B with the equivalent 1 to $N$th hop SINR in the multi-hop system, i.e., $SINR_{1\sim N}^{-1}(k)$, then the previous SVD based power allocation problems in (45) and (51) for the two-hop MIMO relay channel can also be applied to the multi-hop MIMO relay channel and solved based on the GP optimization. For the feasibility test problem, from (66), we have the following optimization problem for the multihop relaying scenario.

$$\min_{p_k, p_k^{r(1)}, \dots, p_k^{r(N)}, \forall k, t} t$$
$$subject\ to$$
$$\gamma_k \left( SINR_{1\sim N}^{-1}(k) \frac{\sum_{i=1, \neq k}^{K} p_i^{r(N)} \left| \hat{g}_{k,i} \right|^2 + \sigma_k^2}{p_k^{r(N)} \left| \hat{g}_{k,k} \right|^2} + SINR_{1\sim N}^{-1}(k) + \frac{\sum_{i=1, \neq k}^{K} p_i^{r(N)} \left| \hat{g}_{k,i} \right|^2 + \sigma_k^2}{p_k^{r(N)} \left| \hat{g}_{k,k} \right|^2} \right) \le t, \ k = 1, \dots, K \tag{67}$$
$$\sum_{k=1}^{K} p_k \le P_b^{\max} \ ,$$
$$\sum_{k=1}^{K} p_k^{r(1)} \le P_r^{\max(1)}, \dots, \sum_{k=1}^{K} p_k^{r(N)} \le P_r^{\max(N)}$$

where $p_k^{r(n)}$, $k=1,\dots,K$, are the downlink power allocation for the $k$th stream at the $n$th RS and and $P_r^{\max(n)}$ denotes the transmit power constraint at the $n$th RS. Similarly, the sum power minimization problem becomes

$$\min_{\mathbf{p}, \mathbf{p}^{r(1)}, \dots, \mathbf{p}^{r(N)}} t$$
$$s.t. \sum_{k=1}^{K} p_k + \sum_{k=1}^{K} p_k^{r(1)} + \dots + \sum_{k=1}^{K} p_k^{r(N)} \le t,$$
$$SINR_{1\sim N}^{-1}(k) \frac{\sum_{i=1, \neq k}^{K} p_i^{r(N)} \left| \hat{g}_{k,i} \right|^2 + \sigma_k^2}{p_k^{r(N)} \left| \hat{g}_{k,k} \right|^2} + SINR_{1\sim N}^{-1}(k) + \frac{\sum_{i=1, \neq k}^{K} p_i^{r(N)} \left| \hat{g}_{k,i} \right|^2 + \sigma_k^2}{p_k^{r(N)} \left| \hat{g}_{k,k} \right|^2} \le \gamma_k, \ k = 1, \dots K \tag{68}$$
$$\sum_{k=1}^{K} p_k \le P_b^{\max} \ ,$$
$$\sum_{k=1}^{K} p_k^{r(1)} \le P_r^{\max(1)}, \dots, \sum_{k=1}^{K} p_k^{r(N)} \le P_r^{\max(N)}.$$

If we combine the multi-hop transmission with subchannel pairing, there will be totally $\left( K! \right)^N$ pairing cases to search from.

Again, we will use the heuristic pairing method but with consideration of the accumulated CINR. That is, at the ($n$-1)th RS, we pair $CINR_{(1)}^{1\sim(n-1)} \geq CINR_{(2)}^{1\sim(n-1)} \geq \cdots \geq CINR_{(K)}^{1\sim(n-1)}$ with $CINR_{(K)}^n \leq CINR_{(K-1)}^n \leq \cdots \leq CINR_{(1)}^n$ for $n = 2, ..., N+1$, where the accumulated CINR, $CINR_{(k)}^{1\sim(n-1)}$, and its ordered subscript are defined similarly as in (65) and Section IV.B.4. The new accumulated CINRs $CINR_{(k)}^{1\sim n}$ are then computed according to (65).

<h2 style="text-align:center">V. NUMERICAL RESULTS</h2>

For the MIMO downlink relay channel in Fig.1, we compare the sum power minimization performance of the two proposed algorithms (simple AF based relay design and SVD based relay design) with the bi-convex design scheme derived in [11] for the cases where all the three algorithms are feasible. The bi-convex scheme iteratively optimizes the precoding matrix at either side (BS or RS) via convex optimization while assuming the precoding matrix of the other side is fixed. Fig. 4(a) and Fig. 4(b) illustrate the minimum sum power required to achieve different target SINR levels for the two and four user cases, respectively. The same SINR target is assumed for all MSs. A simple path loss model where the path loss between nodes $i$ and $j$ depends on the path loss exponent $\eta$ and a reference distance $d_0$, i.e., $P_{loss}^{i,j} = \left( \dfrac{d_{i,j}}{d_0} \right)^{\eta}$ is used. Here, we assume the reference distance is of unit length, i.e., $d_0$ =1 and the value of $\eta$ equals to 4. The distance between the BS and the RS is set to 1/2 and the distance from the RS to the $K$ MSs are set to $d = [d_1, d_2, ..., d_K]$. The initial value of the power allocation matrix $\mathbf{\Lambda}_r$ at the RS is set to an identity matrix $\mathbf{I}_{M_r}$ for the proposed two schemes, and for ease of performance comparison, the initial transfer matrix $\mathbf{Q}$ at the RS for the bi-convex scheme is also set to an identity matrix $\mathbf{I}_{M_r}$. Unless particularly specified, $M_b = M_r = K$ for the following simulations. In Fig. 4(a) and Fig. 4(b), the SVD relay design outperforms the simple AF relay design. The reason is that the power allocation for the MSs at the RS can be jointly optimized in the SVD relay design, as compared to the common power allocation at the RS in the simple AF relay design. Although the bi-convex relay design optimizes the precoding matrices at the BS and the RS individually, its total consumed power is still larger than that of the two proposed schemes in Fig. 4(a) for the two-user case. In the four-user case in Fig. 4(b), it is worse than the SVD relay scheme. This performance gap is because that the bi-convex scheme may fall into local optima without delicate initial setting, and the gap is more pronounced when the differences between the distances between the RS and the MSs are larger. The reason is that the channel attenuation of each RS-MS channel is a random variable with mean $P_{loss}^{RS,MS}$ given by the distance-dependent path loss, and the impact using equal initial power setting at the RS becomes more significant if the channel attenuations are less balanced. In terms of complexity, for the bi-convex scheme, although the solution of the precoding matrix $\mathbf{F}$ at the BS can be obtained by a second order cone programming problem (SOCP) after individual phase adjustment of each column in **F** [11], solving the precoding matrix **Q** at the RS is more involved as **Q** is a common multiplicative factor of all streams. As a result, an approximate scheme to find out the best pre-rotation phase of each user by exhaustive search between 0 to $\pi$ at the RS was proposed [11]. If the phase resolution of each user is $\dfrac{\pi}{L}$, there will be totally $L^K$ +1 SOCP problems (1 for **F** and $L^K$ for **Q**) to be solved for one iteration. Note that the optimal solution of **Q** can be reached reached if $L \to \infty$. $L$ is set to 4 in our simulations. More complexity may be needed to find out the best initial value for the precoding matrix **Q**. For our schemes, only one GP problem is solved. For further comparison of the number of design parameters with the help of off-line optimization tools, since the scheme in [11] was proposed to embed both power allocation and beamforming problem into a SOCP optimization program, the complexity is spread over matrices. While in our schemes, only the power allocation problem resorts to the help of GP while the beamformer weight can be efficiently determined via downlink-uplink duality.

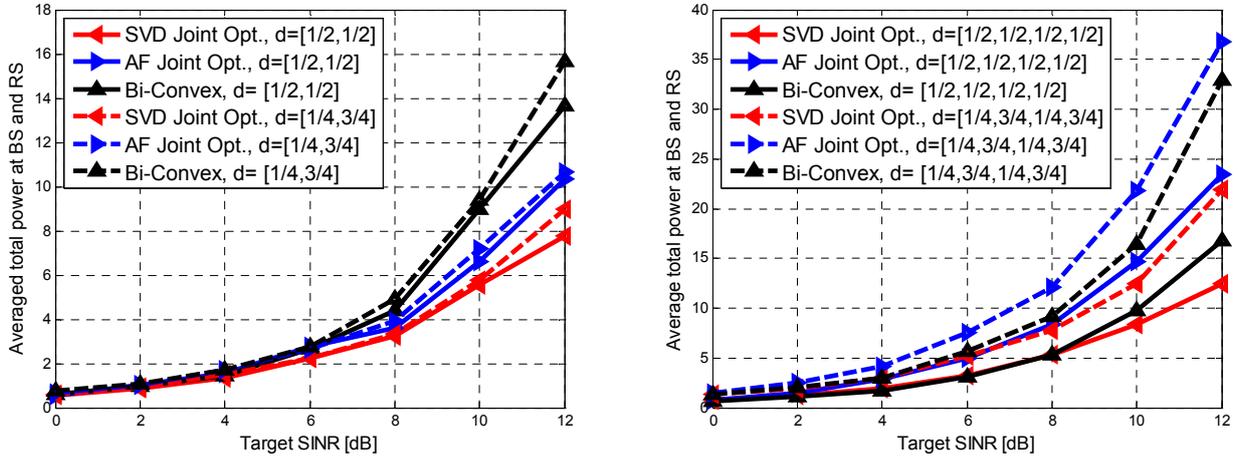

Fig. 4(a). Comparison of the minimum required sum power to achieve different target SINR for $K$=2 users with distances from the RS to the MSs being $d$=[$d1,d2$].

Fig. 4(b). Comparison of the minimum required sum power to achieve different target SINR for $K$=4 users with distances from the RS to the MSs being $d$=[$d1,d2,d3,d4$].

Fig.5 compares the minimal power required as a function of the target SINR and the number of users. It can be observed that the benefit of the SVD based design over the simple AF based design is more apparent with larger numbers of users. This shows that the advantage of the SVD based design, that is, the degree of freedom in power allocation at the RS increases with the number of users.

In Fig.6 we show the average power consumption of the two-user case for the two proposed relay designs as a function of relative distances between the BS, the RS and the MSs. The ratio of the distance from the BS to the RS ($d_{BS\text{-}RS}$) to the distance from the RS to each MS ($d_{RS\text{-}MS}$) is given by $\rho$, i.e., $\rho = d_{BS\text{-}RS}/d_{RS\text{-}MS}$. From Fig. 6, we observe that the performance gap between these two schemes is highly related to the distance ratio. When the RS is closer to the MSs (larger $\rho$), the poor received signal quality at the RS makes the AF based scheme worse than the SVD based scheme due to lack of flexibility to compensate the poor BS-RS channel. When the RS is quite near the BS (smaller $\rho$), the benefit of SVD over AF is less apparent. This is because when the channel between the BS and the RS is much better compared to the channels between the RS and the MSs, the performance is mainly determined by the power allocation at the BS.

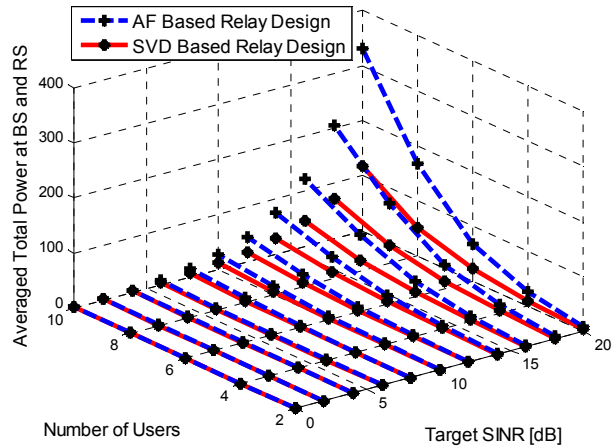

Fig. 5. Total power consumed at the BS and the RS vs. target SINR and the number of users.

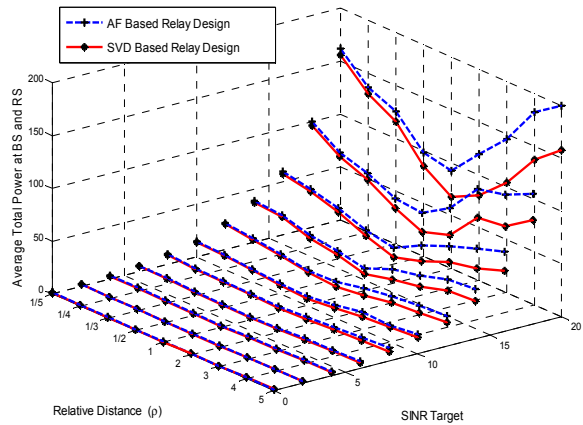

Fig. 6. Average total power consumption vs. distance ratio between two hops and the SINR targets.

In Fig.7, the optimal balanced level is shown for different transmission powers and number of users, where the target SINR is set to 1 for all MSs and the power constraints at the BS and the RS are assumed to be equal with values indicated on the axis. As expected, the SVD relay design generally accommodates more MSs than the AF relay design with the same power constraints. However, if we take a closer look, when the power constraints are low and the number of MSs is high, the AF based design has better balanced levels. The reason is that the SVD scheme may suffer from rank deficiency between the BS-RS link when the number of users increases, while there is no such limitation in the AF scheme.

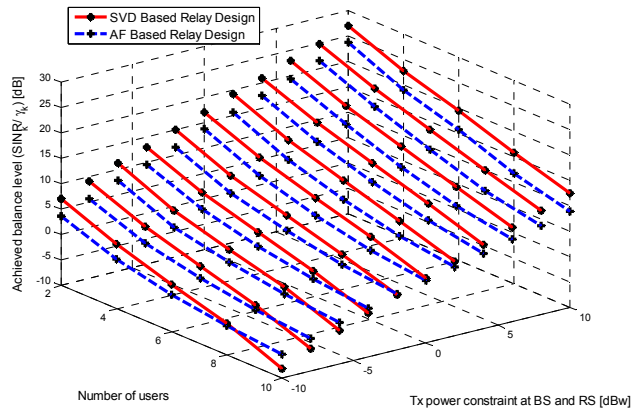

Fig. 7. Achieved balanced level vs. power constraints at the BS and the RS and the number of users.

In Fig.8, we show the convergence speed of the proposed iterative schemes. The total number of iterations (the number of outer loops in Table. I and II) required consists of two stages: the first stage is for the feasibility test and the second stage is for sum power minimization. The transmit power constraints at the BS and the RS are both set to 10w for the two-user case, and the calculation of the average number of iterations is based on the following equation.

$$I_{avg} = \frac{t_{fea}}{t_{total}}\left(I_{test\_fea} + I_{power\_conv}\right) + \frac{\left(t_{total} - t_{fea}\right)}{t_{total}}I_{fea\_conv} \qquad (69)$$

where $t_{total}$ is the total number of channel realizations simulated, $t_{fea}$ is the number of channel realizations that result in feasible results. $I_{test\_fea}$ is the average number of iterations needed to reach a feasible result, i.e., until the balanced level is larger than 1. $I_{power\_conv}$ is the number of iterations needed to have a converged result for the sum power minimization problem and $I_{fea\_conv}$ is the number of iterations needed until the feasibility test converges albeit the result is infeasible. The stopping criterion for convergence is such that the difference of two consecutive iterations is less than 0.001 ($\varepsilon$=0.001). From Fig. 8, the average number of iterations is generally under 5 for different SINR targets. The reason that the number of iterations for SVD based relay design is smaller at higher SINR target is that most of the channel realizations in this region is infeasible and the sum power minimization problem is proceeded less frequently. While for the AF based relay design, there is only one iteration needed for the sum power minimization problem, and the increase of the number of iterations at high SINR is mostly due to the feasibility test problem.

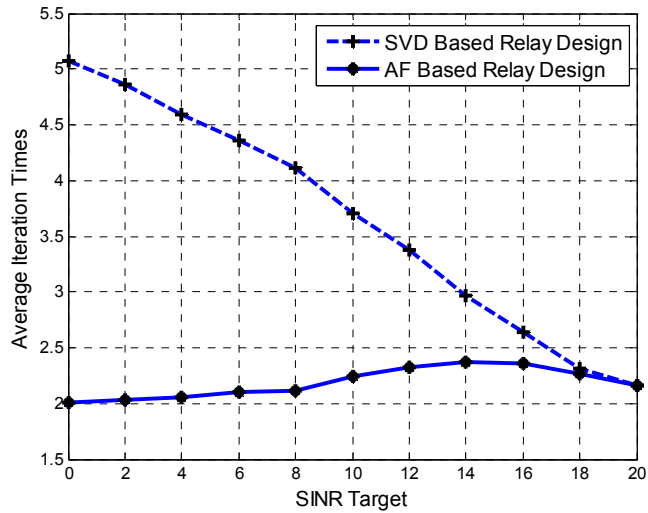

Fig. 8. Average number of iterations needed for the AF and SVD based design with different SINR targets.

In Fig.9, we show the improvement of power efficiency in the SVD based relay design by the subchannel pairing discussed in Section IV.4. We consider two distance settings between the RS and the MSs, denoted as $d = [d1, d2]$ for the two-user case. As we can see, the benefit of subchannel pairing is apparent when the distances to MSs are unbalanced.

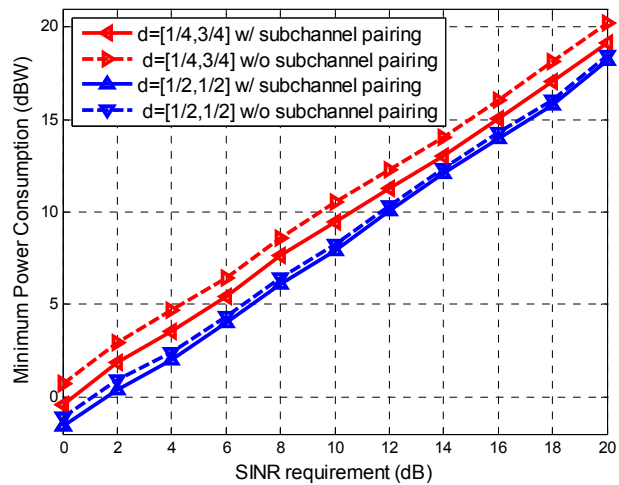

Fig. 9. The comparison of SVD based relay design with and without subchannel pairing for d=[1/2, 1/2] and d=[1/4, 3/4].

To extend the application to the multi-hop MIMO relay in Fig.3 for the SVD based relaying design, a comparison of power efficiency for different numbers of hops is illustrated in Fig.10 which shows the total power consumed at the BS and the RSs under different SINR targets for the number of hops form one to four. In our simulation, the total distance between the BS and the MSs is set to 2, and the RSs are allocated uniformly between the BS and the MSs. The simulation results show that the power efficiency can be improved with more hops. Since the noise accumulated at each RS will be forwarded to the later RSs, the gain by increasing the number of hops saturates. Beyond three hops, the gain is almost not visible. Note that the one-hop model is the MIMO broadcast channel without relay, and can be seen as a special case of our framework.

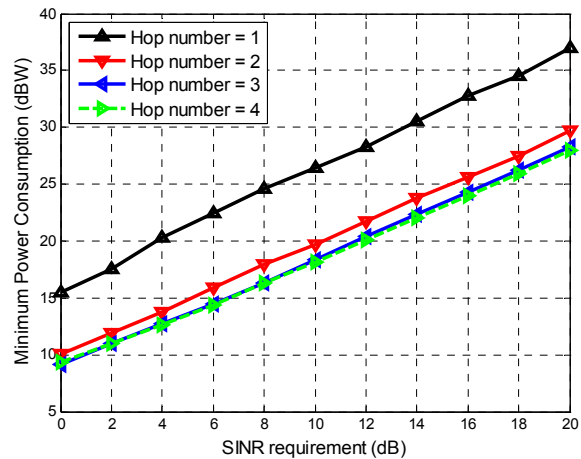

Fig. 10. The comparison of SVD based multi-hop relay design with different numbers of hops.

## VI. CONCLUSION

We proposed joint power allocation and beamforming designs at the BS and the RS via GP optimization and downlink-uplink duality for AF based and SVD based MIMO relay broadcast channels. Iterative algorithms were proposed to iteratively solve three sets of design parameters including the downlink power allocation, virtual uplink power allocation and virtual uplink beamformers for both the feasibility test problem and the power minimization problem. We demonstrated that the proposed SVD relaying scheme has rapid convergence and better power efficiency. Subchannel pairing and multi-hop extension of the SVD based design further improve the power efficiency.

## REFERENCES

[1] I. E. Telatar, "Capacity of multi-antenna Gaussian channels," *European Trans. Tel.*, vol. 10, no. 6, pp. 585-595, Nov. 1999.

[2] G. J. Foschini and M. J. Gans, "On limits of wireless communications in a fading environment when using multiple antennas," *Wireless Pers. Comm.*, vol. 6, no. 3, pp. 311-335, March 1998.

[3] L. Zheng and D. Tse, "Diversity and Multiplexing: A Fundamental Tradeoff in Multiple Antenna Channels", *IEEE Transactions on Information Theory*, vol. 49(5), May 2003.

[4] G. Caire and S. Shamai, "On the achievable throughput of multiantenna Gaussian broadcast channel," *IEEE Trans. Inform. Theory*, vol. 49, no. 7, pp. 1691-1706, Jul. 2003.

[5] A. Goldsmith, S. Jafar, N. Jindal, and S. Vishwanath, "Capacity limits of MIMO channels," *IEEE J. Select. Areas Commun.*, vol. 21, no. 5, pp. 684-702, June 2003.

[6] P. Viswanath and D.N. Tse, "Sum capacity of the vector Gaussian channel and uplink-donwlink duality," *IEEE Trans. Inform. Theory*, vol. 49, no. 8, pp. 1912-192, Aug. 2003.

[7]   C. Chae, T.Tang, R. W. Heath, and S. Cho, "MIMO Relaying With Linear Processing for Multiuser Transmission in Fixed Relay Networks", *IEEE* Trans. Signal Processing, vol. 56, no. 2, pp. 727-738, Feb. 2008.

[8]   S. A. Jafar, K. S. Gomadam, and C. Huang, "Duality and rate optimization for multiple access and broadcast channels with amplify-and-forward relays," *IEEE Trans. Inf. Theory*, vol. 53, no. 10, pp. 3350–3371,Oct. 2007.

[9]   K. S. Gomadam and S. A. Jafar, "On the duality of MIMO MAC and BC with AF relays," in *Proc. IEEE Asilomar Conf. Signals, Syst.Comput.*, Nov. 2007.

[10]  K. T. Phan, T. Le-Ngoc, S. A. Vorobyov, and C. Tellambura,  "Power Allocation in Wireless Multi-User Relay Networks, " *IEEE Trans. Wireless Comm.*, vol. 8, no. 5, pp. 2535-2545, May 2009.

[11]  R. Zhang, C. C. Chai and Y. C. Liang, "Joint Beamforming and Power Control for Multiantenna Relay Broadcast Channel With QoS Constraints," *IEEE Trans. Wireless Comm.,* vol. 57, no. 2,  pp. 726-737, Feb 2009.

[12]  M. Schubert and H. Boche, "Solution of the multiuser downlink beamforming problem with individual SINR constraints," *IEEE Trans. Veh. Technol.,* vol. 53, jo. 1, pp. 18-28, Jan.2004.

[13]  E. Visotsky and U. Madhow, "Optimum beamforming using transmit antenna arrays," in *Proc. IEEE Vehicular Techn. Conf. (VTC) Spring*,vol. 1, Houston, Texas, May 1999, pp. 851–856.

[14]  S. Boyd and L. Vandenberghe, *Convex Optimization*, Cambridge University Press, 2004.

[15]  M. Grant and S. Boyd, CVX: Matlab Software for Disciplined Convex Programming Mar. 2008,  http://stanford.edu/boyd/cvx

[16]  R.A. Horn and C. R. Johnson, Matrix Analysis. Canbridge, U.K.: Canbridge Univ. Press, 1985.

[17]  P. Herhold, E. Zimmermann, G. Fettweis, "Cooperative multi-hop transmission in wireless networks," Computer Networks, Vol. 49, pp. 299-324, 2005.

[18]  J. N. Laneman and G. W. Wornell, "Energy efficient antenna sharing and relaying for wireless networks," *in Pro. IEEE WCNC*, Vol. 1, pp. 23-28 Sept. 2000.